\documentclass[journal]{IEEEtran}
\usepackage{graphicx}
\usepackage{amsmath}
\usepackage{amsfonts}
\usepackage{amssymb, color}

\newtheorem{theorem}{Theorem}
\newtheorem{lemma}{Lemma}
\newtheorem{remark}{Remark}

\begin{document}

\title{Rate Maximization of Decode-and-Forward Relaying Systems with RF Energy Harvesting}
\author{Zoran Hadzi-Velkov, Nikola Zlatanov, Trung Q. Duong, and Robert Schober 
\thanks{Z. Hadzi-Velkov is with the Faculty of Electrical Engineering and Information Technologies, Ss. Cyril and Methodius University, 1000 Skopje, Macedonia (email: zoranhv@feit.ukim.edu.mk).}
\thanks{N. Zlatanov is with the Department of Electrical and Computer Systems Engineering, Monash University, Clayton, VIC 3800, Australia (e-mail: nikola.zlatanov@monash.edu).}
\thanks{T. Q. Duong is with the School of Electronics, Electrical Engineering and Computer Science, Queen's University Belfast, Belfast BT7 1NN, U.K. (e-mail: trung.q.duong@qub.ac.uk).}
\thanks{R. Schober is with the Institute of Digital Communications, Friedrich-Alexander University, Erlangen, Germany (email: robert.schober@fau.de).}
}

\markboth{Accepted for journal publication}{Shell \MakeLowercase{\textit{et al.}}: Bare Demo of IEEEtran.cls for Journals} \maketitle

\begin{abstract}
We consider a three-node decode-and-forward (DF) half-duplex relaying system, where the source first harvests RF energy from the
relay, and then uses this energy to transmit information to the destination via the relay. We assume that the information transfer  and wireless power transfer  phases alternate over time in the same frequency band, and their {\it time fraction} (TF) may change or be fixed from one transmission epoch (fading state) to the next. For this system, we maximize the  achievable average data rate. Thereby, we propose two schemes: (1) jointly optimal power and TF allocation, and (2) optimal power allocation with fixed TF. Due to the small amounts of harvested power at the source, the two schemes achieve similar information rates, but yield significant performance gains compared to a benchmark system with fixed power and fixed TF allocation.

\end{abstract}



\section{Introduction}
Energy harvesting (EH) technology may provide a perpetual power supply to energy-constrained wireless systems, such as  sensor networks. In order to maintain reliable EH-based communication, dedicated far-field radio frequency (RF) radiation may be used as energy supply for the EH transmitters, an approach known as wireless power transfer (WPT) [\ref{lit1}], [\ref{lit2}]. If the signal used for information transfer is also used for simultaneous energy transfer, then a fundamental tradeoff exists between energy and information transfer, as has been shown, e.g. for the noisy channel in [\ref{lit1}], the multi-user channel in [\ref{lit3}], and the multiple-input multiple-output (MIMO) broadcast channel in  [\ref{lit4}]. In [\ref{lit1}] and [\ref{lit3}], the receiver is able to decode information and extract power simultaneously, which, as argued in [\ref{lit4}], may not be possible in practice and the information and energy signals have to be separated  either by time-switching or power-splitting  architectures at the receiver.

EH may also be beneficial in relay networks [\ref{lit5}]-[\ref{lit8}]. In [\ref{lit6}], the authors studied the average data rate of an amplify-and-forward (AF) relaying system for both time-switching and power-splitting protocols. The average data rate and optimal power allocation for decode-and-forward (DF) relaying systems were considered in [\ref{lit7}]. DF relaying with multiple source-destination pairs transmitting at fixed rates was studied in [\ref{lit8}]. In  [\ref{lit5}]-[\ref{lit8}], the relay is an EH node  that harvests RF energy from the source, whereas the source sends both information and energy to the EH relay.

In this paper, we study the performance of a DF relaying system, where the source first harvests RF energy from the relay, and then transmits its information to the destination via the relay. This scenario may arise, for example, in an EH sensor network where the sensor is an EH device that transmits some measured data to a base station via a fixed relay, which is used for  coverage extension and for powering the sensor by WPT. To the best of the authors' knowledge, [\ref{lit9}] is the only other work that considers the case of a relay sending RF energy to an EH source. However, the system model in our paper is somewhat different from the model in [\ref{lit9}], as it assumes: (a) random channel fading, and (b) adaptive time division between the EH and information transmission (IT) phases.


\section{System Model}
We consider a relaying system that consists of an EH source (EHS), a relay, and a destination $D$, where the direct source-destination link is not available. The EHS does not have a permanent power supply, but instead has a rechargeable battery that harvests RF energy broadcasted by the relay. The relay is used both for DF relaying of information from the EHS to $D$, and for broadcasting RF energy to the EHS.

We assume that the source-relay and relay-destination links experience independent fading and independent additive white Gaussian noise (AWGN) with   power $N_0$. The fading in both channels follows the quasi-static (block) fading model, i.e., the channels are constant during each fading block, but change from one block to the next. Let the duration of one fading block be denoted by $T$. In block $i$, the fading power gains of the EHS-relay and relay-$D$ channels are denoted by $x_i'$ and $y_i'$, respectively. For convenience, these gains are normalized by the AWGN power, yielding $x_i = x_i'/N_0$ and $y_i = y_i'/N_0$ with average values $\Omega_X = E[x_i']/N_0$ and $\Omega_Y = E[y_i']/N_0$, respectively, where $E[\cdot]$ denotes expectation. The reverse relay-EHS and $D$-relay channels are assumed to be reciprocal to the respective EHS-relay and relay-$D$ channels.

The information transmission from the EHS to $D$ and the WPT from the relay to the EHS are realized as alternating half-duplex signal transmissions over the same frequency band, organized in transmission epochs. The transmission epochs have a duration of $T$, and therefore, one epoch coincides with the duration of a single fading block. Each epoch consists of three phases: an EH phase, an IT phase 1 (IT1), and an IT phase 2 (IT2). During the EH phase, the relay broadcasts RF energy, which is harvested by the EHS. During IT1, the EHS spends the total energy harvested in the EH phase to transmit information to the relay, which decodes it, and then  forwards this information to $D$ in IT2.

Let us consider $M \to \infty$ transmission epochs. In epoch $i$, the duration of the EH phase is set to $\tau_i T$, and the durations of IT1 and IT2 are both set to $(1-\tau_i)T/2$, where $\tau_i$ is referred to as the {\it time-fraction} (TF) parameter ($0 < \tau_i < 1$). The relay's output power in epoch $i$ is denoted by $p_i$, and assumed constant for the entire epoch duration. In fact, since the relay is primarily a communication device (i.e., its circuitry is not specialized for WPT), the transmit power constraints for the EH and IT phases are similar.

The RF energy received during the EH phase at the source is denoted by $E_{Si}$, and given by $E_{Si} = N_0 p_i x_i \tau_i T$. In IT1, the EHS spends the total harvested energy $E_{Si}$ for transmitting a complex-valued Gaussian codeword of duration $(1-\tau_i) T/2$ with an output power
\begin{equation} \label{rav1}
P_S(i) = \frac{E_{Si}}{(1 - \tau_i) T/2} = \frac{2N_0 p_i x_i \tau_i}{1 - \tau_i}.
\end{equation}
In IT2, the relay forwards all information received from the EHS in IT1 to the destination. Thereby, the relay transmits a   complex-valued Gaussian codeword of length $T(1-\tau_i)/2$ with output power $p_i$. Thereby, the maximum amount of data that can be transmitted successfully from the EHS to the destination via the  relay in epoch $i$ is given by $R(i) = \min \{ C_1(\tau_i, p_i),  C_2(\tau_i, p_i) \}$, where
\begin{equation} \label{c1def}
   C_1(\tau_i, p_i) = \frac{1 - \tau_i}{2} \log \left( 1 + \frac{p_i a_i \tau_i}{1 - \tau_i}   \right),
\end{equation}
\begin{equation} \label{c2def}
 C_2(\tau_i, p_i) = \frac{1 - \tau_i}{2} \log \left( 1 + p _i y_i  \right), \qquad
\end{equation}
with $a_i = 2N_0 x_i^2$. Note that (\ref{c1def}) is obtained by inserting (\ref{rav1}) into $\log(1+P_S(i)x_i)$. As $M\to\infty$, the   achieved average data rate with the proposed scheme is given by
\begin{align}\label{eq_1}
\bar R = \lim\limits_{M\to\infty}\frac{1}{M} \sum_{i=1}^M \min \{ C_1(\tau_i, p_i), C_2(\tau_i, p_i) \}.
\end{align}

\section{Maximization of the Average Data Rate}

We aim at maximizing $\bar R$, given by (\ref{eq_1}), subject to the relay's average power constraint $\bar P$, by proposing two schemes for allocation of the relay's output power $p_i$ and the TF parameter $\tau_i$: (1) the jointly optimal power and time allocation (JOPTA) scheme, and (2) the optimal power allocation (OPA) scheme. In the JOPTA scheme, $p_i$ and $\tau_i$ are jointly optimized, whereas, in the OPA scheme, $p_i$ is optimized for some fixed TF parameter. The data rates achieved with the proposed schemes are feasible if, in each epoch $i$, the three nodes have perfect channel state information (CSI) of both fading links, $x_i$ and $y_i$. These data rates can serve as upper bounds for the imperfect CSI case.

\vspace{-3mm}
\subsection{Jointly Optimal Power and Time Allocation}
We define the following optimization problem for the proposed JOPTA scheme:
\begin{equation} \label{op1}
\underset{p_i\geq 0, \, 0<\tau_i<1, \, \forall i} {\text{max}} \ \frac{1}{M} \sum_{i=1}^M   \min \{ C_1(\tau_i, p_i), C_2(\tau_i, p_i) \} \notag
\end{equation}
\vspace{-4mm}
\begin{eqnarray} \label{rav4}
\text{s.t.} &\text{C1}:& \frac{1}{M} \sum_{i=1}^M p_i \frac{1+\tau_i}{2} \leq \bar P.
\end{eqnarray}
In (\ref{rav4}), $\text{C1}$ represents the relay's average power constraint, which incorporates the relay's power expenditure during all $M$ epochs. The solution of (\ref{rav4}) is given in the following theorem.
\begin{theorem}
The  optimal value of the TF parameter $\tau_i$  is given by
\begin{equation} \label{Sol2}
\tau_i^* = \min \{\tau_{1i}, \tau_{2i} \},
\end{equation}
where
\begin{equation} \label{Sol3}
\tau_{1i} =\left\{
\begin{array}{rl}
\tau_{0i}, & a_i/\lambda > 2
\\
1, & \text{otherwise}
\end{array}
\right. \textrm{\quad and \quad}
\tau_{2i} = \frac{y_i}{a_i + y_i}.
\end{equation}
In (\ref{Sol3}), $\tau_{0i}$ is the root of
\begin{equation}  \label{Sol4}
1 + \frac{a_i \, \tau_{0i}^2}{\lambda \, (1 + \tau_{0i}^2)} \, \log \left[\frac{a_i \, \tau_{0i}}{\lambda \, (1+\tau_{0i})}  \right] = \frac{a_i \, \tau_{0i}}{\lambda \, (1+\tau_{0i})},
\end{equation}
which satisfies $(a_i/\lambda - 1)^{-1} < \tau_{0i} < 1$.
The optimal power allocation at the relay is given by
\begin{equation} \label{Sol1}
p_i^* =\left\{
\begin{array}{rl}
p_{1i}, & \tau_{1i} \leq \tau_{2i}
\\
p_{2i}, & \text{otherwise},
\end{array}
\right.
\end{equation}
where
\begin{equation} \label{Sol6}
p_{1i} = \frac{1-\tau_{1i}}{1+\tau_{1i}} \left(\frac{1}{\lambda} - \frac{1 + \tau_{1i}}{a_i \, \, \tau_{1i}} \right)^+,
\end{equation}
\begin{equation} \label{Sol7}
p_{2i} = \frac{1-\tau_{2i}}{1+\tau_{2i}} \left(\frac{1}{\lambda} - \frac{2}{a_i} - \frac{1}{y_i} \right)^+,
\end{equation}
with $(\cdot)^{+} = \max\{0, \cdot \}$. The constant $\lambda$ is found such that  constraint $\text{C1}$ in (\ref{op1}) holds with equality.
\end{theorem}
\vspace{0mm}
\begin{IEEEproof}
Please refer to Appendix A.
\end{IEEEproof}

\begin{remark}[Roots of (\ref{Sol4})] For $a_i/\lambda > 2$, (\ref{Sol4}) has two roots, denoted by $\tau_{0i}'$ and $\tau_{0i}''$, which satisfy $0 < \tau_{0i}' < \tau_{0i}'' < 1$. The root with the smaller value, $\tau_{0i}'$,  is calculated as $\tau_{0i}' = (a_i/\lambda -1)^{-1}$, but leads to a trivial value for the relay's output power, $p_{i}^* = 0$. The root with the larger value, $\tau_{0i}''$, is the relevant root that leads to $p_{i}^* > 0$.
\end{remark}

\vspace{-3mm}
\subsection{Optimal Power Allocation}
In the OPA scheme, the TF parameter is assumed to have a fixed value, i.e., $\tau_i=\tau_0$,  $\forall i$. Assuming $M\to\infty$, the relay's power allocation that maximizes the average data rate is obtained as the solution of the following optimization problem,
\begin{equation} \label{op2}
\underset{p_i \geq 0, \forall i} {\text{max}} \ \frac{1}{M} \sum_{i=1}^M   \min \left\{ C_1(\tau_0, p_i), C_2(\tau_0, p_i) \right\} \notag
\end{equation}
\vspace{-5mm}
\begin{eqnarray}
\text{s.t.} &\text{C1}:& \frac{1+\tau_0}{2M} \sum_{i=1}^M p_i \leq \bar P.
\end{eqnarray}
The solution of (\ref{op2}) is given in the following theorem.
\begin{theorem}
The relay optimal power allocation is given by
\begin{equation} \label{op2Sol}
p_i^* =\left\{
\begin{array}{rl}
\left(\frac{1}{\lambda} - \frac{1-\tau_0}{a_i \, \tau_0} \right)^+, & \tau_0 \leq \frac{y_i}{a_i + y_i}
\\
\left(\frac{1}{\lambda} - \frac{1}{y_i} \right)^+, & \text{otherwise},
\end{array}
\right.
\end{equation}
where constant $\lambda$ is  found such that $\text{C1}$ in (\ref{op2}) holds with equality.
\end{theorem}
\vspace{0mm}
\begin{IEEEproof}
Please refer to Appendix B.
\end{IEEEproof}
\vspace{+3mm}

Introducing (\ref{op2Sol}) into the objective function of (\ref{op2}), the average data rate $\bar R$ is maximized for given $\bar P$ and $\tau_0$. Let us denote this data rate by $\bar R (\tau_0)$. Clearly, $\tau_0$ should be selected so as to maximize $\bar R (\tau_0)$ for given $\bar P$, as
\begin{equation} \label{op2cmaxmax}
\tau_0^* = \underset{0 < \tau_0 < 1} {\text{arg} \max} \ \bar R(\tau_0).
\end{equation}
Since $\bar R(0) = \bar R(1) = 0$ and $\bar R(\tau_0) > 0$ for $0 < \tau_0 < 1$, according to the Rolle's theorem, there must be a point $\tau_0^*$ between $0$ and $1$ where $\bar R(\tau_0)$ attains a maximum.

The Lagrange multiplier $\lambda$  can be determined by using the following lemma.
\begin{lemma}
Given $\bar P$ and $\tau_0$, the constant $\lambda$ in (\ref{op2Sol}) is determined numerically as the solution of the following expression
\begin{equation}
\frac{2\bar P}{1 + \tau_0} = \int_{\frac{\lambda (1 - \tau_0)}{\tau_0}}^{\infty} \int_{\frac{a \tau_0}{1-\tau_0}}^{\infty} \left(\frac{1}{\lambda} - \frac{1-\tau_0}{a \tau_0} \right) f_Y(y) f_A(a) dy da \notag \\
\end{equation}
\vspace{-1mm}
\begin{equation}\label{lambdafind}
+ \int_{\lambda}^{\infty} \int_{\frac{y (1-\tau_0)}{\tau_0}}^{\infty} \left( \frac{1}{\lambda} - \frac{1}{y}\right) f_A(a) f_Y(y) da dy,
\end{equation}
where $f_A(a)$ and $f_Y(y)$ are the probability density functions (PDFs) of $a_i = 2N_0 x_i^2$ and $y_i$, respectively. Note that $f_A(a)$ can be expressed in terms of the PDF of $x_i$, $f_X(x)$, as $f_A(a) = f_X \big(\sqrt a/(2N_0) \big)/\big(2 \sqrt {2 N_0 a}\big)$.

Once $\lambda$ is determined from (\ref{lambdafind}), the maximum average data rate is calculated as
\begin{eqnarray} \label{op2cmax}
\bar R(\tau_0) = \qquad\qquad\qquad\qquad\qquad\qquad\qquad\qquad\qquad\qquad\quad \notag \\
= \frac{1-\tau_0}{2} \int_{\frac{\lambda (1 - \tau_0)}{\tau_0}}^{\infty} \int_{\frac{a \tau_0}{1-\tau_0}}^{\infty} \log \left(\frac{a\tau_0}{(1-\tau_0) \lambda} \right) f_Y(y) f_A(a) dy da \notag
\end{eqnarray}
\begin{equation}
+ \frac{1-\tau_0}{2} \int_{\lambda}^{\infty} \int_{\frac{y (1-\tau_0)}{\tau_0}}^{\infty} \log \left( \frac{y}{\lambda} \right) f_A(a) f_Y(y) da dy, \quad \,\,
\end{equation}
\end{lemma}
\begin{IEEEproof}
Since $M \to \infty$, the time averages in both $\text{C1}$ and the objective function of (\ref{op2}) are calculated from their corresponding statistical averages, which yields (\ref{lambdafind}) and (\ref{op2cmax}).
\end{IEEEproof}

\vspace{-5mm}
\subsection{Benchmark: Fixed Time and Fixed Power Allocation}
The improvement offered by the two schemes is compared to a fixed power and fixed time allocation (FPTA) scheme. The FPTA scheme assumes fixed relay output power, $P_0$, and a fixed TF parameter, $\tau_0$. For a fair comparison, given the power budget, $\bar P$, for the JOPTA and OPA schemes, the relay's power for the FTPA scheme is fixed to $P_0 = 2 \bar P/(1 + \tau_0) $, whereas $\tau_0$ is set to the same value as the TF parameter of the OPA scheme, calculated based on  (\ref{op2cmaxmax}). In this case, in epoch $i$, the EHS transmits with power $2N_0 P_0 x_i \tau_0/(1 - \tau_0)$. As $M \to \infty$, the achievable rate of the FPTA scheme is determined as
\begin{equation} \label{refrate}
R_0 = \frac{1 - \tau_0}{2 M} \sum_{i=1}^M \log \left(1 + P_0 \cdot \min \left\{y_i, \frac{a_i \tau_0}{1 - \tau_0} \right\} \right).
\end{equation}
In order to achieve $R_0$, both the EHS and relay should transmit with this fixed rate in all epochs by using infinitely-long Gaussian codewords that span infinitely many fading states.

\vspace{-3mm}
\section{Numerical Results}

Fig. 1 illustrates the average achievable data rates of the proposed schemes in block Rayleigh fading. The deterministic path loss is calculated as $E[x_i'] = E[y_i'] = 10^{-4}\, D_i^{-\alpha}$, with the distance between all the nodes set to $D_i = 10$m, the pathloss exponent set to $\alpha = 3$, and the pathloss at a  reference distance of $1$m set to 40dB. We assume an AWGN power spectral density of $-150$dBm/Hz and a bandwidth of $1$MHz, yielding $N_0 = 10^{-12}$ Watts.

Both proposed schemes outperform the FPTA benchmark scheme. The rates achieved by OPA are close to those for JOPTA. However, compared to JOPTA, OPA is simpler and therefore more practical. Namely, the TF parameter of the OPA scheme is fixed for given $\bar P$ (for example, $\tau_0^* = 0.34$ for $\bar P = 1$ Watts, and $\tau_0^* = 0.26$ for $\bar P = 10$ Watts). Additionally, the nodes utilizing OPA can use a single code-book and fixed length codewords, whereas the nodes utilizing JOPTA must change codebooks and the codeword lengths in each transmission epoch.

\begin{figure}[tbp]
\centering
\includegraphics[width=3.7in]{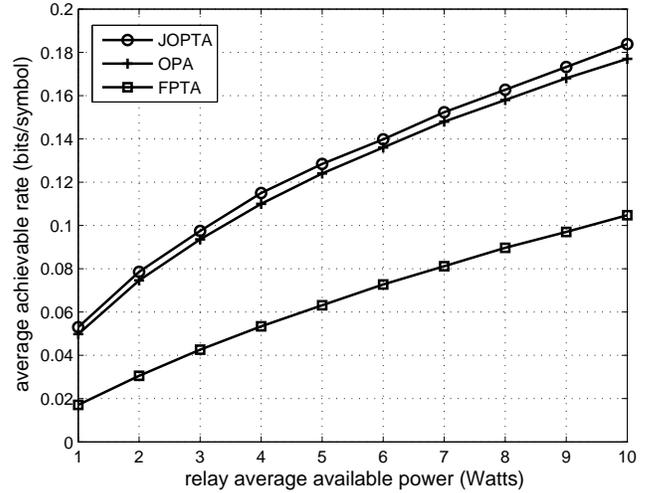} \vspace{-8mm}
\caption{Average achievable rates vs. average relay transmit power} \vspace{-3mm}
\label{fig3}
\end{figure}


\appendices

\vspace{-3mm}
\section{Proof of Theorem 1}

Following [\ref{litNewNew1}, Eqs. (18), (19)], after introducing the change of variables $e_i = p_i(1+\tau_i)/2$, (\ref{op1}) is rewritten as
\begin{equation} \label{op1alt0}
\underset{e_i\geq 0,\; 0<\tau_i<1, \forall i} {\text{max}} \ \frac{1}{M} \sum_{i=1}^M \min \left\{ \bar C_{1}\left(\tau_i, e_i\right), \bar C_{2}\left(\tau_i, e_i\right) \right\} \notag
\end{equation}
\vspace{-5mm}
\begin{eqnarray}
\text{s.t.} &C1:& \frac{1}{M} \sum_{i=1}^M e_i \leq \bar P
\end{eqnarray}
where $\bar C_{1}\left(\tau_{i}, e_i \right) = C_1(\tau_i, 2e_i/(1+\tau_i))$ and $\bar C_{2}\left(\tau_{i}, e_i \right) = C_2(\tau_i, 2e_i/(1+\tau_i))$, with $C_1(\cdot)$ and $C_2(\cdot)$ given by (\ref{c1def}) and (\ref{c2def}), respectively.

Optimization problem (\ref{op1alt0}) is non-convex. Nevertheless, based on [\ref{litNew1}, Theorem 1], we can still apply the Lagrange duality method to solve (\ref{op1alt0}) because of a zero duality gap. In particular, (\ref{op1alt0}) is in the form of [\ref{litNew1}, Eq. (4)]. For any fixed $\tau_i$, $\bar C_{1}(\tau_i,e_i)$ and $\bar C_{2}(\tau_i,e_i)$ are concave in $e_i$. Therefore, for any fixed set of $\tau_i, \forall i$, the objective function of (\ref{op1alt0}) is concave in $(e_1, e_2, ..., e_M)$ for that set of $\tau_i$s, and constraint $C1$ is affine (i.e. convex) in $(e_1, e_2, ..., e_M)$. According to [\ref{litNew1}, Definition 1], the {\it time-sharing} condition is thus satisfied, implying the zero duality gap.

The Lagrangian of (\ref{op1alt0}) is defined as
\begin{equation} \label{LD1}
L\left(e_i, \tau_i, \lambda \right) = \sum_{i=1}^M \min \left\{ \bar C_{1}\left(\tau_i, e_i\right), \bar C_{2}\left(\tau_i, e_i\right) \right\}
- \lambda e_i,
\end{equation}
where $\lambda$ is the Lagrange multiplier associated with C1 of (\ref{op1alt0}). Then, the Lagrangian dual function of (\ref{op1alt0}) is expressed as
\begin{eqnarray} \label{LD2}
L\left(\lambda \right) = \underset{e_i\geq 0,\; 0<\tau_i<1, \forall i} {\text{max}} L\left(e_i, \tau_i, \lambda \right).
\end{eqnarray}
Applying Lagrangian dual decomposition, (\ref{LD2}) can be decoupled into $M$ subproblems (one for each fading state),
\begin{eqnarray} \label{LD3}
\underset{e_i\geq 0,\; 0<\tau_i<1} {\text{max}} \left[ \min \left\{ \bar  C_{1}\left(\tau_i, e_i\right), \bar C_{2}\left(\tau_i, e_i\right) \right\} - \lambda \, e_i \right] \notag \\
= \underset{e_i\geq 0} {\text{max}} \left[H_i(e_i) - \lambda \, e_i \right], \qquad \qquad \qquad
\end{eqnarray}
where \vspace{-3mm}
\begin{eqnarray} \label{Hdef}
H_i(e_i) = \max_{0<\tau_i<1} \, \left\{\min \left[ \bar C_{1}\left(\tau_i, e_i\right), \bar C_{2}\left(\tau_i, e_i\right) \right] \right\}.
\end{eqnarray}

\subsubsection{Solution of (\ref{Hdef})}
For any fixed  $e_i$, we consider two critical points for $\tau_i$: (1) the maximum of $\bar C_{1}(\tau_i, e_i)$, and (2) the intersection between $\bar C_{1}(\tau_i, e_i)$ and $\bar C_{2}(\tau_i, e_i)$. These critical points are denoted by $\bar \tau_{1i}$ and $\bar \tau_{2i}$, respectively.

The {\it critical point (1)} is found from $\partial \bar C_{1}(\tau_i, e_i)/\partial \tau_i = 0$, yielding
\begin{equation} \label{rav5}
\frac{2 e_i a_i (1+\bar \tau_{1i}^2)}{(1 + \bar \tau_{1i}) [1 + (2 e_i a_i  - \bar \tau_{1i}) \bar \tau_{1i}] } = \log \left(1 + \frac{2 e_i a_i \bar \tau_{1i}}{1 - \bar \tau_{1i}^2} \right).
\end{equation}
For any given value of the product $e_ia_i$, $\bar C_{1}(\tau_i, e_i)$ has a single maximum at $\tau_{i} = \bar \tau_{1i}$, which is calculated as the root of (\ref{rav5}). The maximizer $\bar \tau_{1i}$ is a monotonically decreasing function of $e_i$, which satisfies $\bar \tau_{1i} \to 1$ as $e_i \to 0$, and $\bar \tau_{1i} \to 0$ as $e_i \to \infty$.

The {\it critical point (2)} is found from the equality $\bar C_{1}(\bar \tau_{2i}, e_i) = \bar C_{2}(\bar \tau_{2i}, e_i)$, as
\begin{equation} \label{tau2}
\bar \tau_{2i} = \frac{y_i}{a_i + y_i}.
\end{equation}

If $\bar \tau_{1i} \leq \bar \tau_{2i}$, $\min\{$$\bar C_{1}(\bar \tau_{1i}, e_i), \bar C_{2}(\bar \tau_{1i}, e_i)\}$ $=$ $\bar C_{1}(\bar \tau_{1i}, e_i)$ and $H_i(e_i) = \max_{0<\tau_i<1} \{\bar C_{1}(\bar \tau_{1i}, e_i),$ $\bar C_{1}(\bar \tau_{2i}, e_i)\} = \bar C_{1}(\bar \tau_{1i}, e_i)$. If $\bar \tau_{1i} > \bar \tau_{2i}$, $\min\{\bar C_{1}(\bar \tau_{1i}, e_i),$ $\bar C_{2}(\bar \tau_{1i}, e_i)\}$ $=$ $\bar C_{2}(\bar \tau_{1i}, e_i)$ and $H_i(e_i) = \max_{0<\tau_i<1} \{\bar C_{2}(\bar \tau_{1i}, e_i), \bar C_{2}(\bar \tau_{2i}, e_i)\} = \bar C_{2}(\bar \tau_{2i}, e_i)$. Thus, \vspace{-1mm}
\begin{equation} \label{Hdef1}
H_i (e_i) =\left\{
\begin{array}{rl}
U_i(e_i), & \bar \tau_{1i} \leq \bar \tau_{2i}
\\
V_i(e_i), & \bar \tau_{1i} > \bar \tau_{2i}.
\end{array}
\right.
\end{equation}
where $U_i(e_i) \triangleq \bar C_{1}(\bar \tau_{1i}, e_i)$ and $V_i(e_i) \triangleq \bar C_{1}(\bar \tau_{2i}, e_i)$ $=$ $\bar C_{2}(\bar \tau_{2i}, e_i)$.
\subsubsection{Solution of (\ref{LD3})}
Considering (\ref{Hdef1}), we can now solve (\ref{LD3}) by setting the first derivative of $H_i(e_i) - \lambda \, e_i$ w.r.t. $e_i$ to zero. The found critical points are maxima because both $U_i(e_i)$ and $V_i(e_i)$, and therefore $H_i(e_i)$, are concave and monotonically increasing in $e_i$. The concavity of $U_i(e_i)$ and $V_i(e_i)$ is proven from their respective second derivatives w.r.t. $e_i$, which satisfy $U_i''(e_i) < 0$ and $V_i''(e_i) < 0$ for any $e_i > 0$.

Critical point (1): If $\bar \tau_{1i} \leq \bar \tau_{2i} $, (\ref{LD3}) reduces to
\begin{equation} \label{dual1subproblem}
\underset{e_i \geq 0} {\text{max}} \ U_{i}(e_i) - \lambda e_i.
\end{equation}
The solution of (\ref{dual1subproblem}) is found from $U_{i}'(e_i) = \lambda $, where $U_{i}'(e_i)$ denotes the first derivative of $U_i(e_i)$ w.r.t. $e_i$. Since $U_i(e_i)= C_{1}(\bar \tau_{1i}, e_i)$, $U_{i}'(e_i)$ is determined as the total derivative of $C_{1}(\bar \tau_{1i}, e_i)$  w.r.t. $e_i$. At the critical point (1), $\left. \partial\bar C_{1}(\tau_i, e_i) /\partial \tau_i \right|_{\tau_i = \bar \tau_{1i}} = 0$, and, therefore,
\begin{equation} \label{rav11}
U_{i}'(e_i) = \frac{\partial \bar C_{1}\left(\bar \tau_{1i}, e_i\right)}{\partial e_i} = \frac{a_i \, \bar \tau_{1i} (1 - \bar \tau_{1i})}{1 + 2 e_i a_i \bar \tau_{1i} - \bar \tau_{1i}^2}.
\end{equation}
Thus, (\ref{dual1subproblem}) is solved as
\begin{equation} \label{dual1sol}
e_{1i} = \frac{1 - \bar \tau_{1i}}{2} \left(\frac{1}{\lambda} - \frac{1+\bar \tau_{1i}}{a_i \, \bar \tau_{1i}} \right)^+,
\end{equation}
where the $(\cdot)^+$ operator is due to the constraint $e_{i} \geq 0$. Eqs. (\ref{rav5}) and (\ref{dual1sol}) constitute a set of two equations with two unknowns, $e_{1i}$ and $\bar \tau_{1i}$, yielding (\ref{Sol4}) and (\ref{Sol6}).

Critical point (2): If $\bar \tau_{1i} > \bar \tau_{2i} $, (\ref{LD3}) reduces to
\begin{equation} \label{dual2subproblem}
\underset{e_i \geq 0} {\text{max}} \ V_{i}(e_i) - \lambda e_i.
\end{equation}
The solution of (\ref{dual2subproblem}) is the root of $V_{i}'(e_i) = \lambda $ w.r.t $e_i$, yielding
\begin{equation} \label{dual2sol}
e_{2i} = \frac{1 - \bar \tau_{2i}}{2} \left(\frac{1}{\lambda} - \frac{1+\bar \tau_{2i}}{1-\bar \tau_{2i}} \ \frac{1}{y_i} \right)^+,
\end{equation}
where the $(\cdot)^+$ operator is again due to the constraint $e_{i} \geq 0$. Combining (\ref{tau2}) and (\ref{dual2sol}), we obtain (\ref{Sol7}).

\vspace{-2mm}
\section{Proof of Theorem 2}

The convex optimization problem (\ref{op2}) is rewritten as
\begin{equation}
\underset{p_i \geq 0} {\text{max}} \  \frac{1}{M} \sum_{i=1}^M \, H_i(p_i), \quad \text{s. t.} \quad \frac{1}{M} \sum_{i=1}^M p_i \leq 2\bar P/(1+\tau_0),
\end{equation}
where
\begin{eqnarray} \label{op2dual}
H_i(p_i) &=& \min \left\{ C_{1}(\tau_0, p_i), C_{2}(\tau_0, p_i) \right\} \notag \\
&=&\left\{
\begin{array}{rl}
C_{1}(\tau_0, p_i), & y_i \geq a_i \tau_0/(1-\tau_0)
\\
C_{2}(\tau_0, p_i), & \text{otherwise}.
\end{array}
\right.
\end{eqnarray}
Decomposing Lagrangian dual function into $M$ subproblems (one for each fading state), we obtain $\max_{p_i \geq 0} H_i(p_i) - \lambda p_i$, which is solved as the root of $H'(p_i) = \lambda$ w.r.t. $p_i$.

\end{document}